\documentclass[margin=1inch]{article}
\usepackage[textwidth=8em,textsize=small]{todonotes}
\usepackage{natbib}

\usepackage{amsmath,amsfonts,amssymb,amsthm,epsfig, graphicx, hyperref, mathtools, comment, caption, subcaption, float,enumitem}
\usepackage{amsmath,amsfonts,amssymb,amsthm, mathtools, float, adjustbox, graphicx,algorithm2e}
\newcommand{\beq}{\begin{equation}}
\newcommand{\eeq}{\end{equation}}
\newcommand{\beas}{\begin{align*}}
\newcommand{\eeas}{\end{align*}}
\newcommand{\bea}{\begin{align}}
\newcommand{\eea}{\end{align}}
\newcommand{\bei}{\begin{itemize}}
	\newcommand{\eei}{\end{itemize}}
\newcommand{\ben}{\begin{enumerate}}
	\newcommand{\een}{\end{enumerate}}
\newcommand{\bet}{\begin{theorem}}
	\newcommand{\eet}{\end{theorem}}
\newcommand{\bel}{\begin{lemma}}
	\newcommand{\eel}{\end{lemma}}
\newcommand{\bep}{\begin{proposition}}
	\newcommand{\eep}{\end{proposition}}
\newcommand{\bed}{\begin{definition}}
	\newcommand{\eed}{\end{definition}}
\newcommand{\bec}{\begin{corollary}}
	\newcommand{\eec}{\end{corollary}}
\newcommand{\bex}{\begin{example}}
	\newcommand{\eex}{\end{example}}

\newcommand{\bu}{\mathbf{u}}
\newcommand{\bw}{\mathbf{w}}

\newcommand{\bs}{\mathbf{s}}

\newcommand{\bW}{\mathbf{W}}

\newcommand{\bR}{\mathbf{R}}

\newcommand{\bY}{\mathbf{Y}}
\newcommand{\by}{\mathbf{y}}
\newcommand{\bX}{\mathbf{X}}
\newcommand{\bx}{\mathbf{x}}

\newcommand{\bSig}{\boldsymbol{\Sigma}}

\newcommand{\bdelta}{\boldsymbol{\delta}}

\newcommand{\bbeta}{\boldsymbol{\beta}}

\newcommand{\btheta}{\boldsymbol{\theta}}

\newcommand{\R}{\mathbb{R}}
\newcommand{\E}{\mathbb{E}}

\newcommand{\argmin}{\mathop{\rm arg\min}}

\usepackage{multirow}
\usepackage[utf8]{inputenc}

\newcommand{\vertiii}[1]{{\left\vert\kern-0.25ex\left\vert\kern-0.25ex\left\vert #1 
		\right\vert\kern-0.25ex\right\vert\kern-0.25ex\right\vert}}

\newtheorem{theorem}{Theorem}
\newtheorem{corollary}{Corollary}

\newtheorem{assumption}{Assumption}

\title{\textbf{\Large Robust angle-based transfer learning in high dimensions}}
\author{\large Tian Gu$^{1}$, Yi Han$^{2}$, and Rui Duan$^{3,\dag}$}
\date{\large $^1$ Department of Biostatistics, Columbia Mailman School of Public Health, New York, NY 10032, USA\\
$^2$ Department of Statistics, Columbia University, New York, NY 10027, USA\\
$^3$Department of Biostatistics, Harvard T.H. Chan School of Public Health, \\
Boston, MA 02115, USA\\
$^\dag$ Corresponding author: rduan@hsph.harvard.edu}

\begin{document}
\maketitle
\begin{abstract}
\normalsize
Transfer learning enhances the performance of a target model by leveraging data from related source populations, a technique particularly beneficial when target data is scarce. This study addresses the challenge of training high-dimensional regression models with limited target data, in the context of heterogeneous source populations.  We consider a practical setting where only the parameter estimates of the pre-trained source models are accessible, instead of the individual-level source data. Under the setting with only one source model, we propose a novel flexible angle-based transfer learning (angleTL) method, which leverages the concordance between the source and the target model parameters. We showed that the proposed angleTL is adaptive to the signal strength of the target model, unifies several benchmark methods by construction, and can prevent negative transfer when between-population heterogeneity is large. We also provide algorithms to effectively incorporate multiple source models accounting for the fact that some source models may be more helpful than others. Our high-dimensional asymptotic analysis provides interpretations and insights on when a source model can be useful for the target, and demonstrates the superiority of angleTL over other benchmark methods. We perform extensive simulation studies to validate our theoretical conclusions and show the feasibility of applying angleTL to transferring existing genetic risk prediction models across multiple biobanks.
\end{abstract}
\newpage

\section{\Large Introduction}

Insufficient training data presents a critical challenge across various domains. In finance, the scarcity of comprehensive user credit histories hampers efforts in evaluating individual financial risk and detecting fraud \citep{teja2017indonesian}. Similarly, in precision medicine, the limited availability of medical records, especially for studying rare diseases or minority and disadvantaged sub-populations, compromises both the fairness and clinical efficacy of treatments \citep{jia2017towards, kim2019minority}. These constraints highlight an urgent need for innovative approaches that can enhance model performance in the face of data limitations \citep{Viele2014, Li2020, YangKim2020}.

When borrowing information from external sources, a major challenge is to account for the potential data heterogeneity. The external data may likely be collected from a different population with diverse characteristics \citep{chen2020propensity} or be historical data where the variable definitions and measures may change over time \citep{mansukhani2019effect, mitchell2021polygenic}. Many data integration methods are proposed to leverage potentially shared information across populations while addressing data heterogeneity. For example, some methods are based on re-weighting or re-sampling the source data such that they are more similar to the target \citep{huang2006correcting, pan2010domain, long2013transfer}; some methods assume there is a unique lower dimensional representation of features across populations, which can be transferred from the source to the target \citep{tzeng2014deep, ganin2015unsupervised, sun2016deep}; some methods propose to use the target data to calibrate the source models \citep{girshick2014rich,li2020transfer,gu2022transRF}. {In situations where no data from the target population is available, methods have been proposed to combine source models  aiming for distributional robustness by optimizing for the worst-case performance, assuming the target population can be represented as either a single source or a mixture of source populations \citep{maximin,wang2023distributionally}.} The performance of many of the aforementioned methods largely depends on whether the underlying assumptions regarding the similarity between the source and the target populations hold, which is usually unknown in practice. Therefore, it is desired to develop methods that can be adaptive to the underlying data heterogeneity or at least prevent the case where incorporating the source information leads to worse model performance than not including it, known as the ``negative transfer" phenomenon \citep{weiss2016survey}.

In addition, there might be data-sharing constraints  such that external sources  cannot share individual-level data with the target study. Federated or distributed algorithms are proposed to overcome such data-sharing barriers by sharing only summary-level statistics across studies, many of which rely on sharing the gradients or higher-order derivatives of objective functions \citep{ duan2018odal, duan2019ODAC, duan2022heterogeneity,cai2021individual,li2021targeting} and may require iteratively sharing updated summary statistics across datasets \citep{li2021targeting}. However, distributed and federated learning are less feasible without a collaborative environment or certain infrastructures that enable efficient computing and timely information sharing. In contrast, pre-trained models from existing studies are often more accessible. With increasing attention to reproducibility and open science, more journals require studies to publish their  results as supplementary materials or to make them shareable upon request \citep{thompson2006assessing, roobol2012prediction}. Many platforms allow direct implementation \citep{belbasis2022reproducibility} or validation of fitted models on secure collaborative platforms such as PheKB \citep{kirby2016phekb} and FATE \citep{liu2021fate}. There is an increasing need for data integration methods that can directly leverage fitted models to improve the model performance of a target study.

Regression models are broadly applied in many fields, due to advantages such as simplicity, computational efficiency, and interpretability \citep{hafemeister2019normalization,wynants2020prediction,wu2020exposure}, and they are also the building blocks of many data analysis pipelines and tools \citep{VanBuuren2006,tan2006regression}. In this paper, we consider the problem of incorporating pre-trained regression models from external sources to help train  a target  model with limited target data. For the $i$-th subject in the target data, let $Y_i\in \R$ denote the outcome variable of interest and $\bX_i\in \R^p$ denote a set of $p$-dimensional covariates. We consider
\[
Y_i = \bX_i^\top\bbeta + \epsilon_i,\quad \text{for } i \in \{1, \dots, n\}
\]
where $\epsilon_i \in \R$ is the random noise with mean zero and variance $\sigma^2$, $\bbeta \in \R^p$ is the regression coefficient of interest, and $n$ is the target sample size. In addition to the target data, we observe a source model fitted on an external source dataset, where we observe the source estimate $\hat{\bw}\in \R^p$ of the underlying regression coefficient $\bw$ in the source population. In the case where the underlying coefficients $\bbeta$ and $\bw$ show certain similarities, we hope to leverage $\hat{\bw}$ to assist the estimation of the target parameter $\bbeta$.

Similar problems have been considered in the literature. In a series of transfer learning work, the source model estimate $\hat\bw$ was incorporated through a regularization with the form $\|\bbeta-\hat\bw\|_q$, for some positive constant $q$. For example, \citet{li2020transfer} proposed the transLASSO algorithm that leverages the source data by adding a penalty term $\|\bbeta-\hat\bw\|_1$ when learning $\bbeta$ using the target data. Later, transLASSO has been extended to generalized linear models \citep{Tian2022transGLM}, functional linear regression \citep{lin2022transfer}, and Q-learning \citep{chen2022transferred}. Considering data sharing constraints, \citet{li2021targeting} proposed a federated learning approach by sharing gradients and Hessian matrices, and \citet{gu2022commute} proposed a method to incorporate information from fitted models through a synthetic data approach. Similar ideas are also used in multi-tasking learning literature, where $L_2$-norm-based penalties, i.e., $\|\bbeta-\bw\|_2$, are introduced to leverage the similarities between model parameters  \citep{tian2022unsupervised,duan2022adaptive}. The  underlying similarity assumption of these methods is that the $L_q$ distance between $\bbeta$ and $\bw$ is small, and we refer to this class of methods as the distance-based transfer learning methods.

In real-world applications, the distance-based transfer learning methods may be less effective when $\bbeta$ and $\bw$ are highly concordant, but their distance may not be small. For example, the source outcome may be defined differently from the target outcome (categorical versus continuous characterization of the same outcome), or there might be standardization procedures applied to the source data that are unknown to us.  The source model might be fitted using a different but related outcome variable  correlated with the target outcome \citep{miglioretti2003latent, stearns2010one}. To leverage the concordance between the model parameters, several alternative similarity characterizations are proposed. For example, \citet{Zhao2014BivariateRidge} proposed a bivariate ridge regression assuming the $j$-th entries $(\beta_j, w_j)$ follows a bivariate normal distribution with shared correlation $\rho$ for all $j \in [p]$. Similarly, \citet{Ripke2015BLUP} proposed a multi-trait prediction method, assuming multivariate normally distributed random effects to leverage shared genetic architectures across traits. \citet{qiao2019decomposition} proposed a method for multi-objective optimization problems. \citet{liang2020learning} proposed a calibration version of transLASSO, which allows the scale of the model parameters to differ. However, these methods require individual-level data from the source and their robustness is unclear when the level of data heterogeneity is high.

{In this paper, we propose an \textbf{angle}-based \textbf{T}ransfer \textbf{L}earning approach, named angleTL, which leverages the similarity of two regression models through a novel penalization obtained by decoupling the angle distance between $\bbeta$ and $\bw$.  As a consequence, angleTL adapts to signal strength and inherently guards against negative transfer, offering a simpler form both conceptually and computationally. Our approach unifies distance-based transfer learning,  target-only and source-only estimators as specific instances, and thus is guaranteed to have superior performance. The high-dimensional asymptotic analysis provides a precise characterization of the prediction risk, illustrating the bias-variance trade-off, the influence of signal strengths, the similarity between source and target, the noise level, and the estimation error of the source to the accuracy of the final estimator. Given multiple source models, we propose methods to effectively incorporate them toward better predictive performance in the target population. The proposed methods only require parameter estimates of fitted source models, which are more accessible in practice.  We perform extensive simulation studies to validate our theoretical conclusions and evaluate the performance of angleTL by training genetic risk models for low-density lipoprotein cholesterol (LDL) using data from multiple large-scale biobanks.}

\section{\Large Angle-based transfer learning} \label{method}
Let $\bY\in \R^{n}$ denote the outcome variable of interest and $\bX \in \R^{n\times p}$ denote a set of $p$-dimensional covariates in the target data of size $n$. Without borrowing information from the source,  we can obtain a \it{target-only estimator} \rm of $\bbeta$ through a ridge regression, 
\begin{align} \label{ridge_tradition}
\begin{split} 
    \tilde \bbeta_\lambda &= \argmin_{\bbeta} \frac{1}{n}\|\bY - \bX\bbeta\|_2^2 +\lambda\|\bbeta\|_2^2,
\end{split}
\end{align}
where $\lambda$ is a tuning parameter. The  penalty on $\|\bbeta\|_2^2$ helps reduce the overall mean squared error of the estimator in the high-dimensional setting \citep{hastie2009elements}.

\begin{figure}
\centering
\includegraphics[width=1\textwidth]{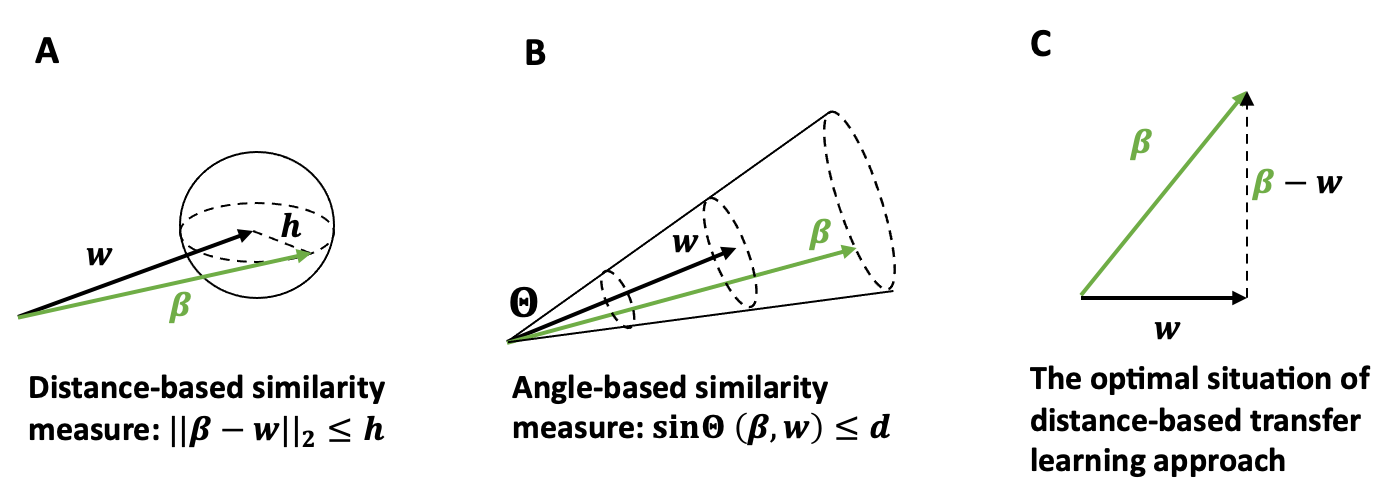} 
\caption{Geometric illustration of the distance-based similarity characterization (A); the angle-based characterization (B); and the situation where the distance-based transfer learning have the same predictive risk as the proposed method (C).} \label{similarity_measure}
\end{figure}

Suppose that we also observe $\hat{\bw}\in \R^p$, the parameter estimates of a source model fitted on an external source dataset with a sample size potentially much larger than $n$. Due to data heterogeneity, the two underlying regression coefficients, $\bbeta$ and $\bw$, may not be the same but may share certain similarities, and hence $\hat{\bw}$ can be used to guide the estimation of $\bbeta$. Following a series of recent transfer learning methods \citep{li2022transfer, li2021targeting,gu2022commute}, we define the \it{ $L_2$-distance-based transfer learning estimator} \rm (distTL) to be
\begin{align} \label{distance-based_solution}
\begin{split}
    \check \bbeta_{\lambda_d} &= \argmin_{\bbeta} \frac{1}{n}\|\bY - \bX \bbeta\|_2^2 +\lambda_d\|\bbeta-\hat{\bw}\|_2^2, 
\end{split}
\end{align}
where $\lambda_d$ is a tuning parameter. Imposing a distance-based penalty, $\|\bbeta - \hat\bw\|_2$, is equivalent to adding a constraint $\|\bbeta - \hat\bw\|_2\le h$, which reduces the parameter space to an $L_2$ ball centered at $\hat\bw$ as shown in panel A of Fig. \ref{similarity_measure}.  
Anchoring on the source estimator $\hat \bw$, distTL encourages estimators closer to $\hat\bw$ while allowing for the calibration of potential differences. 



As introduced earlier, in real-world applications, it is possible that $\bbeta$ and $\bw$ are concordant to some degree, but $\|\bbeta - \bw\|_2$ is not small. In such cases, distTL may be less effective and we may consider a more general characterization of the similarity using the angle distance characterized by $\sin\Theta (\bw, \bbeta) = \sqrt{1-\frac{({\bw}^\top\bbeta)^2}{\|{\bw}\|_2^2\|\bbeta\|_2^2}}$ . As shown in panel B of Fig.\ref{similarity_measure}, if the target model parameter $\bbeta$ is restricted by a constraint that $\sin\Theta (\bbeta, \bw)\le d$, the source model parameter $\bw$ can provide directional information of $\bbeta$ and reduce the parameter space of $\bbeta$ to a cone. When $\bw$ and $\bbeta$ are small in $L_2$ distance, it also implies that the angle between the two vectors is small, while it may not be true the other way around.

{Instead of directly penalizing on $\sin\Theta (\hat\bw, \bbeta) = \sqrt{1-\frac{(\hat{\bw}^\top\bbeta)^2}{\|\hat{\bw}\|_2^2\|\bbeta\|_2^2}}$ which may lead to computational complexity, we consider an alternative \emph{angle-based transfer learning estimator} defined as
\begin{align} \label{proposed_optimization}
\begin{split}
    \hat \bbeta_{\lambda,\eta} 
    &= \argmin_{\bbeta} \frac{1}{n}\|\bY - \bX \bbeta\|_2^2 +\lambda\|\bbeta\|_2^2 -2\eta\hat\bw^\top\bbeta,
\end{split}
\end{align}
where ${\lambda}$ and ${\eta}$ are tuning parameters. To see the connection between the proposed penalty terms and the $\sin\Theta$ distance, without loss of generality, we can consider $\|\hat\bw\|_2^2=1$. Due to the duality of constrained optimization and penalization, the penalty on $\sin\Theta (\hat\bw, \bbeta)$ is equivalent to having a constraint $\sqrt{1-\frac{(\hat{\bw}^\top\bbeta)^2}{\|\bbeta\|_2^2}}< a$ for some $a$, which can be written  as $\frac{\hat{\bw}^\top\bbeta}{\|\bbeta\|_2}>\sqrt{1-a^2}$. While this constraint controls the term $\frac{\hat{\bw}^\top\bbeta}{\|\bbeta\|_2}$, what we propose is to decouple its numerator and denominator, and introducing two separate constraints  $\hat{\bw}^\top\bbeta>b$ (equivalent to $-\hat{\bw}^\top\bbeta<-b$) and $\|\bbeta\|_2<c$ (equivalent to  $\|\bbeta\|_2^2<c^2$). While this decoupling provides computational simplicity and efficiency, the set ${ \bbeta: \hat{\bw}^\top\bbeta > b, \|\bbeta\|_2 < c}$ may not necessarily include, or be included in, the set ${\bbeta: \sqrt{1-\frac{(\hat{\bw}^\top\bbeta)^2}{\|\bbeta\|_2^2}} < a}$,  when $b$ and $c$ are two freely adjustable tuning parameters. So the proposed penalization is not equivalent to or is a relaxation of the $\sin\Theta$ penalty.}

{
The proposed penalization has several advantages. 
First, given $\lambda$ and $\eta$, Equation (\ref{proposed_optimization}) has a closed-form solution $\hat \bbeta_{\lambda,\eta}=(\bX^\top \bX+n \lambda \mathbf{I}_p)^{-1}(\bX^\top \bY+n \eta \hat{\bw})$, which ensures computational efficiency. Secondly, it is adaptive to the signal strength of the target parameter. Specifically, when $\|\bbeta\|_2$ is large, the constraint $\hat\bw^\top\bbeta>b$ can be satisfied for $\bbeta$'s with large $\sin\Theta$ distance with $\hat\bw$. In this scenario, the angle constraint is relatively minimal, and less information is borrowed from the source. Conversely, when $\|\bbeta\|_2$ is small, there is a tighter constraint on the angle distance, which results in borrowing more information from the source. Intuitively, with a sufficiently strong signal, relying more on target data can yield precise estimations. However, in situations with weak signal strength, and given that the target sample size might be limited, there is a greater need to leverage information from the source. The adaptivity to the signal strength is verified in our theoretical analysis discussed in Section \ref{theory}. On the contrary, the penalty $\sqrt{1-\frac{(\hat{\bw}^\top\bbeta)^2}{\|\bbeta\|_2^2}}$ takes the same value for all $\bbeta$'s  having the same $\sin\Theta$ distance with $\hat\bw$  regardless of the signal strength $\|\bbeta\|_2$.}  
Thirdly, although our proposed penalty is termed as angle-based penalty, it allows borrowing information from the source through both the angle similarity and the $L_2$ similarity, whereas the $\sin\Theta$ penalty only focuses on the angle similarity. To see this, the penalty terms we proposed can be written equivalently as the following
\[
\lambda \|\bbeta\|_2^2 -2\eta \bbeta^\top\hat{\bw} = \lambda \|\bbeta-\hat\bw\|_2^2 - 2(\eta-\lambda)\bbeta^\top\hat\bw-\lambda\|\hat\bw\|_2^2
\]
where the third term on the right-hand side will not influence the choice of $\bbeta$ in the optimization. We denote the first two terms as $\mathcal P_1(\bbeta;\hat \bw) = \lambda \|\bbeta-\hat\bw\|_2^2$ and $\mathcal P_2(\bbeta;\hat \bw) = 2(\eta-\lambda)\bbeta^\top\hat\bw$.  With fixed signal strength, the second term $\mathcal P_2$  encourages the estimator of $\bbeta$ to be more concordant to $\hat \bw$,  allowing information from the source estimator $\hat \bw$ through the angle similarity; With a fixed angle distance,  $\mathcal P_1$ will then encourage the estimator of $\bbeta$ to be closer to $\bw$ in $L_2$ distance, allowing borrow information from the $L_2$ similarity. Moreover, angleTL incorporates several estimators as special cases: (i) when $\eta = 0$, $\hat \bbeta_{\lambda,\eta}$ reduces to the target-only estimator shown in Equation (\ref{ridge_tradition}), (ii) when $\eta = \lambda$, $\hat \bbeta_{\lambda,\eta}$ becomes distTL in Equation (\ref{distance-based_solution}), since the optimization remains the same if we add an additional $\bbeta$-independent term $\lambda\|\hat \bw\|_2^2$, i.e., $\lambda\|\bbeta\|_2^2 -2\lambda \hat \bw^\top\bbeta +\lambda\|\hat \bw\|_2^2 = \lambda \|\bbeta - \hat \bw\|_2^2$; and (iii) if $\lambda$ and $\eta$ are both extremely large, the penalty term $-2\eta \hat\bw^\top\bbeta$ will dominate and result in $\bbeta \approx c\hat\bw$, the same as rescaling the source estimator by a constant $c$.

Using cross-validation, we can select the optimal $\lambda$ and $\eta$ from a grid covering a proper range. As one of the main advantages of the proposed method, in light of the above discussions, we expect that under suitable conditions, the numerical performance of angleTL will be no worse than the target-only, source-rescaling, or distTL. In particular, the guaranteed performance no worse than the target-only method implies that our method automatically prevents negative transfer without needing a validation dataset, which has great benefits in practice. 

From the perspective of bias-variance trade-off, we see that when $\lambda = \eta = 0$, our estimator $\hat \bbeta_{\lambda,\eta}$ reduces to the ordinary least square estimator, which is an unbiased estimator of $\bbeta$. On the other extreme, if $\lambda$ goes to infinity faster than $\eta$, $\hat \bbeta_{\lambda,\eta}$ converges to $0$, which has large bias and small variance. The contribution of $\eta$ to the bias-variance trade-off depends on the underlying similarity between $\bw$ and $\bbeta$. When $\hat\bw = \bbeta$, if choosing $\eta = \lambda$ and letting $\eta\rightarrow \infty$, $\hat \bbeta_{\lambda,\eta}$ converges to $\bbeta$. In more realistic cases when $\hat\bw \ne \bbeta$, there exists an optimal $\eta$ and $\lambda$ which minimize the mean square error (MSE). In Section \ref{theory}, we will elaborate on these points under a rigorous theoretical framework, and provide precise characterizations of angleTL's performance, revealing its scope of applicability and the advantages over alternative approaches.

\section{\Large Predictive risk and theoretical justifications} \label{theory}

In this section, we study the predictive risk of $\hat \bbeta_{\lambda,\eta}$ in Equation (\ref{proposed_optimization}). For a given choice of $\lambda$ and $\eta$, we denote the prediction risk by
\[
r_{\lambda, \eta} (\bX) = \E\{(\bx^\top\hat\bbeta_{\lambda,\eta} - \by)^2|\bX \}
\]
where $\bX$ is the training data, 
and the expectation is taken over an independent test data point ($\bx, \by$), from the
same distribution as the training data. Given $\lambda$ and $\eta$, we denote $\mathcal{R}{(\lambda, \eta)} = \lim_{(n,p)\to\infty} r_{\lambda, \eta} (\bX)$, the limiting prediction risk as the training sample size $n$ and the dimension $p$ go to infinity. Our theoretical analysis is based on the following assumptions.

\begin{assumption} \label{HDA assumption}
\textbf{(High-Dimensional Asymptotics)} 
\begin{enumerate}
\item [(a)]  The  dimension of the model, $p$, goes to infinity as $n$ goes to infinity,  with $\frac{p}{n} \rightarrow \gamma \in (0, \infty)$.

    \item[(b)]  The covariate variables of each observation in the target population are identically and independently distributed  with mean zero and covariance matrix $\bSig\in\R^{p\times p}$, which is positive semidefinite.
    \item [(c)]  The cumulative distribution function of the eigenvalues, i.e., the spectral distribution, $F_{\bSig}$ of $\bSig$ converges to a limit population spectral distribution on the support $[0, \infty)$.
\end{enumerate}
\end{assumption}

\begin{assumption} \label{RRC assumption}
\textbf{(Random Regression Coefficients)}
The target regression coefficients,
$\bbeta\in \R^{p}$, are random with $\E [\bbeta] = \mathbf{0}$, and $Var[\bbeta] = \frac{\alpha_t^2}{p}{\bf I}_p$. The source coefficient $\bw$ are random with $\E [\bw] = \mathbf{0}$ and $Var[\bw] = \frac{\alpha_s^2}{p}{\bf I}_p$. 
The coefficients $\bbeta$ and $\bw$ satisfy $Cov(\bbeta, \bw) = { \frac{\rho\alpha_t\alpha_s}{p}}{\bf I}_p$.
\end{assumption}

\begin{assumption} \label{IEE assumption}
\textbf{(Source Estimation Error)}
The obtained source estimator $\hat \bw$ can be decomposed as $\hat \bw = \bw + \bdelta$, where $\bdelta$ is some sub-Gaussian estimation error vector independent of $\bw$ and $\bbeta$, with $\E[\bdelta\bdelta^{\top}] =\frac{1}{p} \bSig_{\delta}$. We assume that as $(n,p)\to\infty$, the eigenvalues of $\bSig_{\delta}$ is supported on $[C_L, C_U]$ for some constants $C_U \ge C_L \ge 0$.
\end{assumption}

A few remarks on the above assumptions are in order. In general, these assumptions together help us leverage results from random matrix theory 
to compute the prediction risk of the proposed method, and enable rigorous comparisons with alternative methods. In particular, Assumption \ref{HDA assumption}(a) concerns the limit of the ratio $\frac{p}{n}$ as $n\to\infty$, reflecting the high-dimensional nature of the problem. Assumption \ref{HDA assumption}(b) is a mild condition on the design matrix, which requires independent samples but allows general correlations among high-dimensional covariates. Assumption \ref{HDA assumption}(c) ensures the existence of proper limit for the eigenvalues of the underlying covariance matrix $\bSig$ as $p\to\infty$, which can be satisfied by a large class of covariance matrices with diagonal, exchangeable, or auto-regressive structures \citep{bai2010spectral}. {Assumption \ref{RRC assumption} is a random-effect hypothesis assuming that the true effect size of each feature is drawn independently at random, and the overall signal strength captured in the target (source) model is characterized by the quantity $\alpha_t^2$ ($\alpha_s^2$), where  $\E \|\bbeta\|_2^2=\alpha_t^2$ and $\E \|\bw\|_2^2=\alpha_s^2$.  This hypothesis enables us to study the precise predictive risk for angleTL, distTL, and target-only estimator, and draw clear insight on factors influencing their relative performance. Such a random-effect assumption is commonly seen in high-dimensional statistics and applications to genetic risk prediction \citep{dobriban2018high,Zhao2014BivariateRidge,vilhjalmsson2015modeling}. Finally, Assumption 3 characterizes the estimation error of the source estimate $\hat\bw$ for estimating $\bw$. Depending on the correlation structure of the features, as well as the method used to obtain $\hat\bw$, the estimation error $\bdelta$ may be biased or correlated, and we quantify the magnitude of the error by the upper and lower bounds of eigenvalues, i.e., $C_L$ and $C_U$. While these assumptions facilitate the theoretical analysis, our method can be applied to various other settings. }


Before stating our main theorem, we define a complex function $v(z): \mathbb{C}\setminus \R^+ \to \mathbb{C}$, which plays an important role in our description of the prediction risk $\mathcal{R}(\lambda,\eta)$. Specifically, according to Assumption \ref{HDA assumption}, the spectral distribution of the $n\times n$ matrix $\frac{1}{n}\bX\bX^{\top}$ has a limit, which we denote as $F_{\underline{\bSig}}$. In particular, $F_{\underline{\bSig}}$ is related to $F_{\bSig}$ in Assumption \ref{HDA assumption}(c) via the equation
\[
F_{\underline{\bSig}}(x)-\gamma F_{\bSig}(x)=(1-\gamma)1\{x=0\}.
\]
We define $v(z)$ to be the Stieltjes transform of the probability measure $F_{\underline{\bSig}}$, i.e.,
\[
v(z):= \int_0^\infty \frac{dF_{\underline{\bSig}}(x)}{x-z},\qquad z\in \mathbb{C}\setminus \R^+.
\] 

\begin{theorem}\label{withnoise} 
	Under Assumptions 1-3, as $n\to \infty$, the limiting expected predictive risk $\mathcal{R} (\lambda, \eta)$  satisfies 
	\[
R(\lambda, \eta, C_L)	\le \mathcal{R} (\lambda, \eta) \le R(\lambda, \eta, C_U),
	\]
	where 
	\[
R(\lambda, \eta, C) := 	({\lambda^2\alpha_t^2+\eta^2\alpha_s^2  +\eta^2C - 2\lambda\eta\rho\alpha_t\alpha_s})\frac{v(-\lambda)-\lambda v'(-\lambda)}{\gamma(\lambda v(-\lambda))^2}+\frac{\sigma^2v'(-\lambda)}{v(-\lambda)}.
\]  
The minimal risk $\mathcal{R}^*=\min_{\lambda,\eta}\mathcal{R}(\lambda,\eta)$ under the optimal choice of tuning parameters satisfies 
\[
\frac{\sigma^2}{\lambda_L^* v(-\lambda_L^*)}  \le \mathcal{R}^*\le \frac{\sigma^2}{\lambda_U^* v(-\lambda_U^*)},
\] 
where $\lambda_L^* = \frac{\gamma\sigma^2}{\alpha_t^{2}(1-\frac{\rho^2\alpha_s^2}{\alpha_s^2+C_L})}$ and  $\lambda_U^* = \frac{\gamma\sigma^2}{\alpha_t^{2}(1-\frac{\rho^2\alpha_s^2}{\alpha_s^2+C_U})}$. 
\end{theorem}

The proof of the above theorem can be found in Appendix 1 of the Supplementary Material, which relies on the asymptotic theory of eigenvalues of large random matrices and in particular the theoretical framework developed by \citep{dobriban2018high} for the analysis of ridge-penalized linear regression and linear discrimination analysis. Similar techniques have been used for the precise analysis of many other high-dimensional problems such as covariance matrix estimation \citep{ledoit2015spectrum, cai2020limiting}, low-rank matrix denoising \citep{donoho2014minimax}, sketching and random projection \citep{yang2021reduce}, etc.

Theorem \ref{withnoise} provides upper and lower bounds for the limiting prediction risk $\mathcal{R}(\lambda,\eta)$ as a function of the parameters $\lambda$ and $\eta$. The possible range of $\mathcal{R}(\lambda,\eta)$ is rooted in the variability of the source estimator $\hat\bw$. More specifically, the upper and lower bounds for $\mathcal{R}(\lambda,\eta)$ are determined by the spectrum of $\bSig_\delta$ in Assumption \ref{IEE assumption}, which in general decreases as the estimation error $\bdelta$ becomes smaller. In addition, the difference between the upper and the lower bounds $R(\lambda, \eta, C_U)-R(\lambda, \eta, C_L)=\frac{(C_U-C_L)\eta^2[v(-\lambda)-\lambda v'(-\lambda)]}{\gamma(\lambda v(-\lambda))^2}$ depends on the spectral range $(C_U-C_L)$ of $\bSig_\delta$, which becomes smaller when the components of $\bdelta$ are less heteroskedastic or less correlated.  For example, if $\bSig_\delta = I_p$, we have $C_U=C_L$.

Moreover, the minimal risk $\mathcal{R}^*$ lies in the interval $[\frac{\sigma^2}{\lambda_L^* v(-\lambda_L^*)}, \frac{\sigma^2}{\lambda_U^* v(-\lambda_U^*)}]$. To better understand its implications, we consider a few special scenarios of interest, and assume for simplicity that the population spectral distribution $F_{\bSig}$ is supported on a set bounded away from zero and infinity and $C_L=C_U=C$. In this case, our predictive risk is precise with $
\mathcal{R}^* = \frac{\sigma^2}{\lambda^* v(-\lambda^*)}
$
where $\lambda^* = \frac{\gamma\sigma^2}{\alpha_t^{2}(1-\frac{\rho^2\alpha_s^2}{\alpha_s^2+C})}$ and $\eta^* = \lambda^*\frac{\rho\alpha_t\alpha_s}{\alpha_s^2+C}$.
On the one hand, when the target signal strength is very large, i.e., $\alpha_t^2\to\infty$,  we have that 
\beq \label{rate1}
\mathcal{R}^* \to \frac{\sigma^2}{1-\gamma},\qquad \text{when $\gamma<1$}
\eeq
 and
\beq \label{rate2}
\mathcal{R}^*= \frac{\alpha_t^2(1-\frac{\rho^2\alpha_s^2}{\alpha_s^2+C})}{\gamma v(0)} (1+o(1)),\qquad \text{when $\gamma>1$}.
\eeq
When $\gamma<1$ ($p<n$) and $\alpha_t^2\to\infty$, the prediction risk reduces to that of the ordinary least squares, and is independent of the covariance $\bSig$ and the source data (such as $\rho$, $\alpha_s^2$, or $C$). When $\gamma>1$ ($p>n$), the limiting error as $\alpha_t^2\to\infty$ depends on the covariance $\bSig$ through $v(0)$, and the source estimator through $1-\frac{\rho^2\alpha_s^2}{\alpha_s^2+C}$. In the special case where $\bSig={\bf I}_p$, we have $v(0)=\frac{1}{\gamma-1}$, so that the prediction risk decreases when the sample size increases, when the source-target similarity $\rho^2$ increases, or when the relative source estimation error $\frac{C}{\alpha_s^2}$ decreases. On the other hand, when the target signal strength is very small, i.e., $\alpha_t^2\to 0$, we have 
\beq \label{rate3}
\mathcal{R}^* = \sigma^2+\sigma^2\alpha_t^{2}(1-\frac{\rho^2\alpha_s^2}{\alpha_s^2+C}) T(1+o(1)),
\eeq
where $T=\lim_{n\to\infty}\frac{1}{p}\text{tr}(\bSig)$, so that the difficulty of the prediction is determined to the first order by the average variance of the covariates and the source data, which no longer depends on $\gamma$.

Another scenario of interest is when the source data has strong signal-to-noise ratio or contains extremely large sample size such that the estimation error of source model is negligible, i.e., $\frac{\alpha_s^2}{\alpha_s^2+C_U}\to 1$. 
Under the optimal choice of tuning parameters  
$
\lambda^* = \frac{\gamma\sigma^2}{\alpha_t^{2}(1-\rho^2)}
$
and $\eta^* =\lambda^*\rho\frac{\alpha_t}{\alpha_s}$, we obtain $\mathcal{R}_0^* = \frac{\sigma^2}{\lambda^* v(-\lambda^*)}.$ 
In particular, in the case with no estimation error, i.e., $C_U = 0$, the limiting risk is
\begin{equation}\label{eq:noiseless}
 \mathcal{R}(\lambda, \eta)
=({\lambda^2\alpha_t^2+\eta^2\alpha_s^2  - 2\lambda\eta\rho\alpha_t\alpha_s})\frac{v(-\lambda)-\lambda v'(-\lambda)}{\gamma(\lambda v(-\lambda))^2}+\sigma^2\frac{v'(-\lambda)}{v(-\lambda)}.
\end{equation}
The proofs of Equations (\ref{rate1}) to (\ref{eq:noiseless}) can be found in the Supplementary Material.

The significance of Theorem \ref{withnoise} also lies in its implications on the practical benefits of the proposed method. On the one hand, it suggests that angleTL can automatically avoid negative transfer due to imprecise or low-quality source estimators, with performance no worse than the target-only estimator. Specifically, the limiting predictive risk of the target-only estimator $\tilde \bbeta_\lambda$ has been obtained in \cite{dobriban2018high}, which has a minimum risk
$
\mathcal{R}^\dagger = \frac{\sigma^2}{\lambda^\dagger v(-\lambda^\dagger)}
$
under the optimal choice of tuning parameters $\lambda^\dagger = \gamma\frac{\sigma^2}{\alpha_t^{2}}$. As a comparison, the minimal risk $\mathcal{R}^*$ of the angleTL estimator is always smaller than $\mathcal{R}^\dagger$. To see this, we first notice that $\mathcal{R}^*$ is bounded by $R_U^*=\frac{\sigma^2}{\lambda_U^* v(-\lambda_U^*)}$. Since $\frac{1}{\lambda v(-\lambda)}$ is monotonically decreasing with $\lambda$, and that $\lambda^\dagger =\gamma\frac{\sigma^2}{\alpha_t^{2}}\le\lambda^* =\frac{\gamma\sigma^2}{\alpha_t^{2}(1-\frac{\rho^2\alpha_s^2}{\alpha_s^2+C_U})}$ for any $\rho\ge 0$, $\alpha_s^2\ge 0$ and $C_U>0$, we have that $\mathcal{R}^*\le \mathcal{R}_U^* \le \mathcal{R}^\dagger$. This indicates that the angleTL estimator safeguards against negative transfer because angleTL is always no worse than the target-only estimator, even when source has a large distance to the target (i.e., $\rho$ is small), or the estimation error is large (i.e., $\frac{\alpha_s^2}{\alpha_s^2+C_U}$ is small). The same property is  achieved in \cite{li2020transfer} and \cite{li2021targeting} using an independent validation dataset, which is not required in our method. 

Furthermore, angleTL is expected to have better performance than distTL. Specifically, since distTL is obtained by setting $\lambda = \eta$, under the current framework, its predictive risk is higher than the predictive risk of angleTL, as $\mathcal{R}^*$ is optimized in a two dimensional space that covers $\lambda = \eta$. In the extreme case where $\hat\bw=\bw$, we see that $\lambda^* = \eta^*$ if and only if $\rho \frac{\alpha_t}{\alpha_s} = 1$. Since $\alpha_t^2 = \E\|\bbeta\|^2_2$ and $\alpha_s^2=\E\|\bw\|^2_2$, $\alpha_t$ and $\alpha_s$ can be viewed as the vector lengths of $\bbeta$ and $\bw$, respectively. Intuitively, one would think that distTL requires the lengths of two vectors to be roughly the same, or $\alpha_t = \alpha_s$, to achieve the good performance. However, the optimal choices of $\lambda$ and $\eta$ demonstrates that this is not necessarily the case. Since $\rho$ is the cosine of the angle between the two vectors, it implies that distTL is optimal if and only if the difference vector $\bbeta-\bw$ is orthogonal to $\bw$, as shown in panel C of Fig. \ref{similarity_measure}. This is a rather strong assumption which may  be rarely satisfied in real applications. In contrast, the proposed angleTL depends only on the angle between the source and target parameters. This explains the flexibility and the potential advantage of our angle-based method.

\section{\Large Incorporating multiple source models} \label{multi_source}
In this section, we consider the setting where multiple source models are available, denoted as $\{\hat \bw_k\}_{k=1,\hdots,K}$, and we want to incorporate these source models when fitting a model using the internal target data. In the existence of multiple sources, we hope that our estimator obtained from transfer learning is robust to the cases where some source models are less helpful than others. 

One direct extension of angleTL is to include multiple penalty terms, one for each source estimates $\hat \bw_k$, and obtain
\begin{equation*}
    \hat \bbeta_{\lambda,\eta_1\dots,\eta_K } = \argmin_{\bbeta} \frac{1}{n}\|\bY - \bX \bbeta\|_2^2 +\lambda\|\bbeta\|_2^2 - 2\sum_{k=1}^K \eta_k \hat \bw_k^\top \bbeta,
\end{equation*}
which consists of $K+1$ tuning parameters, $\lambda$ and $\{\eta_k\}_{k=1,\hdots,K}$. However, this might be computationally less feasible when $K$ is large, since optimization involving a large number of tuning parameters will be challenging in terms of computational speed and numerical stability. {Intuitively, if we can derive an aggregated source estimator, denoted as \(\hat{\bw}\), that resides within the convex cone formed by the source models, i.e., $\hat{\bw}\in\{\sum_{i=1}^K\theta_k\hat{\bw}_k: \theta_k\ge0\}$, and aligns more closely with the target parameter \(\bbeta\) than any individual \(\hat{\bw}_k\), then we can directly utilize the single source {angleTL} method detailed in Section \ref{method}.}

Following this idea, we present in this section two approaches to obtain the aggregated estimator $\hat \bw$ from the $K$ source estimators $\{\hat \bw_k\}_{k=1,\hdots,K}$, where the main difference between them lies in whether an independent validation dataset in addition to the training data is needed for the aggregation step. In the existence of an independent validation dataset, we can use it to learn the weights $\btheta = (\theta_1, \dots, \theta_k)$ that linearly combine $\{\hat \bw_k\}_{k=1,\hdots,K}$ to obtain the aggregated estimator $\hat\bw$. This can be achieved via methods such as the Q-aggregation, summarized in Algorithm \ref{agg1} below, which is shown to approximate the best linear combination of all models when the validation data is sufficiently large \citep{RT11, Tsybakov14, Qagg}. {If we have evidence where some of the source models can be highly different from the target, we can further impose sparse structures on $\btheta$.}

\RestyleAlgo{ruled}
\begin{algorithm}[hbt!]
	\caption{Obtain $\hat \bw$ from a validation dataset}\label{agg1}
	\KwData{Source estimates $\{\hat \bw_k\}_{k=1,\dots,K}$ and a validation dataset $\{\mathring{\bX}, \mathring{\bY}\}$}
	Obtain the weights $\boldsymbol{\hat{\theta}}=(\hat\theta_1,\dots,\hat\theta_K)^\top$ by:
	$\boldsymbol{\hat{\theta}} = \argmin_{\theta}{||\mathring{\bY}-\mathring{\bX} (\sum_{k=1}^K\theta_k \hat\bw_k})||_2^2 +\lambda_\theta \mathcal{P}(\theta)$,
 subject to $\btheta\ge 0$, where $\lambda_\theta \mathcal{P}(\theta)$ can be added to enforce sparsity or other desirable structures.\\
	Obtain $\hat \bw = \sum_{k=1}^K\hat{\mathbf{\theta}}_k \hat\bw_k$.\\
	\KwResult{$\hat \bw$}
\end{algorithm}

Alternatively, one could consider using the following spectral approach, summarized in Algorithm \ref{agg2}, to obtain an aggregated $\hat \bw$ without a validation dataset.  Specifically, we first normalize each $\hat\bw_k$ to obtain $\bar\bw_k = \frac{\hat\bw_k}{\|\hat\bw_k\|_2}$, and define $\bar \bW = [\bar \bw_1^\top,\bar \bw_2^\top,\hdots,\bar \bw_K^\top]^\top\in\R^{K\times p}$. Let $\bu_1$ be the first eigenvector of $\bar \bW \bar \bW^\top\in \R^{K \times K}$, and define $\hat \bs = (\hat s_1,...,\hat s_K)= |\bu_1|$, the absolute value of $\bu_1$. We propose to use $\hat \bs$ as the weights for aggregating the $K$ normalized source estimates. Then the final adaptive weighted source estimates is defined as $\hat \bw = \sum_{k=1}^K\hat{{s}}_k \bar\bw_k$. Intuitively, this approach can be considered as carrying out a principal component (PC) analysis on the matrix $\bar\bW$ that combines all the source estimators, where $\bu_1$ is the first PC loadings of $\bar\bW$, under which the linear combination $\bu_1^\top\bar\bW$ of $\hat\bw_k$'s  has the largest variance, or summarizes the most information in $\{\hat\bw_k\}_{k =1,\dots,K}$. In particular, under suitable conditions (see Theorem \ref{stochastic.s.thm} below) on the overall quality of the source estimators, the first PC loadings are all nonnegative (in this case $\hat \bs = \bu_1$), so that the first PC, $\bar\bW^\top\hat \bs$, is exactly the final aggregated source estimator $\hat \bw$. Moreover, whenever $\{\hat\bw_k\}_{k =1,\dots,K}$ together contain sufficient amount of information about the direction of $\bbeta$, it can also be shown that the components of $\hat\bs$ reflect the true discrepancy between each $\hat\bw_k$ and $\bbeta$, and that the final aggregated source estimator is asymptotically no worse than the best candidate source estimator (see Theorem \ref{stochastic.s.thm} below). Similar spectral weighting idea has been considered for combining multiple classifiers without labeled data \citep{parisi2014ranking}. Such a strategy avoids splitting extra samples from already limited target data for aggregation so that it does not affect the prediction performance. 


\RestyleAlgo{ruled}
\begin{algorithm}[hbt!]
\caption{Obtain $\hat \bw$ without a validation dataset}\label{agg2}
\KwData{Source estimates $\{\hat \bw_k\}_{k=1,\dots,K}$}
\For{$k=1,\dots,K$}{
Normalize $\hat{\bw}_k$ and obtain $\bar \bw_k=\frac{\hat \bw_k}{\|\hat{\bw}_k\|_2}$.
}
Define $\bar \bW = [\bar \bw_1^\top,\bar \bw_2^\top,\hdots,\bar \bw_K^\top]\top$. Let $\hat \bs = (\hat s_1,...,\hat s_K) = |\bu_1|$, where $\bu_1$ is the first eigenvector of $\bar \bW \bar \bW^\top$. \\
Obtain the aggregated source estimates $\hat \bw = \sum_{k=1}^K\hat{s}_k \bar\bw_k$.\\
\KwResult{$\hat \bw$}
\end{algorithm}

The theoretical guarantee for Algorithm \ref{agg1} has been carefully studied, e.g., in \cite{Tsybakov14}. Here, we provide theoretical justification for Algorithm \ref{agg2}. We start with some definitions. Similarly as in Section \ref{method}, for each $k \in [K]$, we consider $\hat\bw_k = \bw_k +\bdelta_k$, and assume $\bw_k$ and $\bdelta_k$ satisfy Assumptions 2 and 3. Moreover, we define $\bR_{\bw}=(\rho_{ij}^{\bw})\in\R^{K\times K}$, where  $\frac{\rho_{ij}^{\bw}}{p}$ is the largest eigenvalue of 
\beq
\E\bigg[\bigg(\frac{\bw_i}{\|\bw_i\|_2}-\frac{\bbeta}{\|\bbeta\|_2}\bigg)\bigg(\frac{\bw_j}{\|\bw_j\|_2}-\frac{\bbeta}{\|\bbeta\|_2}\bigg)^\top\bigg], \qquad 1\le i,j\le K,
\eeq
which characterizes the pairwise covariance between the discrepancies in $\frac{\bw_i}{\|\bw_i\|_2}$ and $\frac{\bw_j}{\|\bw_j\|_2}$ with respect to $\frac{\bbeta}{\|\bbeta\|_2}$,
and define $\bR_{\bdelta}=(\rho_{ij}^{\bdelta})\in \R^{K\times K}$ where $\frac{\rho_{ij}^{\bdelta}}{p}$ is the largest eigenvalue of
\beq
\E \bigg[\frac{\bdelta_i\bdelta_j^\top}{\|\bw_i\|_2\|\bw_j\|_2}\bigg],\qquad 1\le i,j\le K
\eeq
which characterizes the  pairwise covariance between the normalized source estimation errors $\frac{\bdelta_i}{\|\bw_i\|_2}$ and $\frac{\bdelta_j}{\|\bw_j\|_2}$.
In a special case where the source datasets are mutually independent, we have $\rho_{ij}^{\bdelta}=0$ for $i\ne j$, and $\bR_{\bdelta}$ is a diagonal matrix. Finally, we define the true similarity between $\hat\bw_i$'s and $\bbeta$ as $\bs=(\cos\angle(\hat\bw_1,\bbeta), ..., \cos\angle(\hat\bw_K,\bbeta))^\top.$
With the above preparation, we state our main results concerning Algorithm \ref{agg2}.

\bet \label{stochastic.s.thm}
Suppose for each $k\in\{1,\dots,K\}$, $\bw_k$ and $\bdelta_k$ satisfy Assumptions \ref{RRC assumption} and \ref{IEE assumption} with positive $\rho$ being bounded away from 0, and that $\|\bdelta_k\|_2^2 = \E\|\bdelta_k\|_2^2 \cdot (1+o(1))$ with high probability\footnote{An even $A_n$ holds with high probability if there  exists some $N>0$ such that for all $n\ge N$ we have $P(A_n)\ge 1-n^{-D}$  for some large constant $D>0$.}. Then, for any small constant $\epsilon>0$, there exists some sufficiently small constant $C_1(\epsilon)>0$ such that whenever $\max\{\|\bR_{\bdelta}\|,\|\bR_{\bw}\|\}/K\le C_1(\epsilon)$, the following holds:
\begin{enumerate}
	\item For the spectral weights $\hat\bs$ and the consensus source estimator $\hat\bw$ defined in Algorithm \ref{agg2}, it holds that $
\cos \angle(\widehat\bs, \bs)=\frac{\widehat\bs^\top \bs}{\|\widehat\bs\|_2\|\bs\|_2}\ge 1-\epsilon$ and $\cos\angle(\hat \bw,\bbeta)>\rho-\epsilon$ in probability as $n\to\infty$. 
     \item There exists some constant $C_2(\epsilon)>0$ such that, whenever $\frac{1}{p}\text{tr}(\bSig_{\delta})\ge C_2(\epsilon)\alpha_s^2$, we have $\cos\angle(\hat \bw,\bbeta)>\max_{1\le k\le K}\cos \angle(\widehat\bw_k,\bbeta) +\epsilon$ in probability as $n\to\infty$.
 \end{enumerate}
\eet

{
Theorem \ref{stochastic.s.thm} yields the effectiveness of the spectral weighting approach in three aspects. Firstly, it ensures that the estimated weight vector $\hat \bs$ will converge to the true similarity measure $\bs=(\cos\angle(\hat\bw_1,\bbeta), ..., \cos\angle(\hat\bw_K,\bbeta))^\top$ as $n\to\infty$. As a result, more weights will be given to the source estimators closer to $\bbeta$ and less weights to those with smaller $\cos\angle (\hat\bw_k,\bbeta)$. Secondly, it indicates the guaranteed performance of the final consensus source estimator $\hat\bw$, as long as the original correlation $\rho$ between the source and target regression coefficients is sufficiently strong. Thirdly,  and interestingly, part two of Theorem \ref{stochastic.s.thm} shows that,  whenever the source estimators have weak signal-to-noise ratio in the sense that  $\frac{\E \|\bw_k\|^2_2}{\E \|\bdelta_k\|_2^2} = \frac{p\alpha_s^2}{\text{tr}(\bSig_{\delta})}$ is small, the consensus estimator has strictly better performance than all the source estimators in the large sample limit, demonstrating the consensus power of the spectral weighted estimator. 

The validity of the method relies on the additional condition that $\max\{\|\bR_{\bdelta}\|,\|\bR_{\bw}\|\}$ is small compared with $K$, which essentially requires the source estimators to be sufficiently diverse and less correlated in terms of both the respective true coefficients $\bw_k$, and the estimation errors $\bdelta_k$. {Specifically, from our proof of Theorem 2 in Appendix 2 of the Supplementary Material, it can be seen as an ``approximate rank-one" assumption for the matrix $\bar{\bf W}$ of normalized source estimators $\{\bar{\bf w}_k\}$. As such, the validity of Theorem 2 would effectively rely on the ``spectral gap" between the singular values of the matrix $\bar{\bf W}$, which can be accessed empirically. In other words, when the ratio between the first two singular values is significantly larger than those between any other consecutive singular values, the assumptions behind Algorithm 2 and Theorem 2 are likely satisfied. In this connection, several data-driven methods such as ScreeNot \citep{donoho2023screenot} and BEMA \citep{ke2023estimation} can be adopted to help assess the appropriateness of this ``approximate rank-one" assumption. Second, due to the robustness of our transfer learning estimator against negative transfer, even when multiple sources are inappropriately aggregated for some reason, the final estimator would not be driven astray. Third, in addition to our spectral approach, one can alternatively use the supervised proposed in Algorithm \ref{agg1}, when the validity of Algorithm \ref{agg2} is unclear.  In Section \ref{realdata}, we show the effectiveness and superior performance of such spectral weighting approach in combining multiple source estimators learned from different genetic risk models.}}

\section{\Large Simulation study} \label{simulation}
\subsection{Comparing empirical and theoretical predictive risks}

We perform a simulation study to verify the theoretical predictive risks shown in Theorem \ref{withnoise}.  For $j\in \{1, \dots, p\}$, we generate 
\begin{equation}\label{simugen}
(\bw_j, \bbeta_j)^\top \overset{i.i.d}{\sim} N\left(0,\frac{1}{p}\begin{pmatrix}\alpha_s^2& \rho\alpha_s\alpha_t\\\rho\alpha_s\alpha_t&\alpha_t^2\end{pmatrix}\right).
\end{equation}
We set $n=50$ and $p=\gamma n$, and generate $\bX$ from independent standard normal distributions. We then generate $\bY \sim N(\bX\bbeta, 0.5\mathbf{I}_n)$ given $\bX$. We vary the correlation between $\bbeta$ and $\bw$, i.e., $\rho \in \{0.3,0.6,0.9\}$, and the signal strength ratio $\frac{\alpha_t}{\alpha_s} \in \{\frac{10}{9}, 2\}$,  as well as the dimension-to-sample ratio $\gamma = \frac{p}{n} \in \{2, \frac{1}{2}\}$.

We first consider the noiseless case where the source estimator $\hat\bw = \bw$. In the noiseless case, we have a precise prediction risk as shown in Equation (\ref{eq:noiseless}). For a given $\lambda$ chosen from a grid, we set $\eta=\rho\frac{\alpha_t}{\alpha_s}$, which is the optimal choice of $\eta$, and calculate the theoretical risk. We also obtain the theoretical risk of distTL by setting $\eta = \lambda$ in Equation (\ref{eq:noiseless}). For the target-only approach, we obtain the theoretical risk by setting $\eta = 0$ in Equation (\ref{eq:noiseless}). To obtain the empirical predictive risks, we apply each method on the training data for each $\lambda$ in the grid, and then obtain the corresponding mean square errors evaluated on an independent testing dataset of size $100$.

\begin{figure}
\centering
\includegraphics[width=\textwidth]{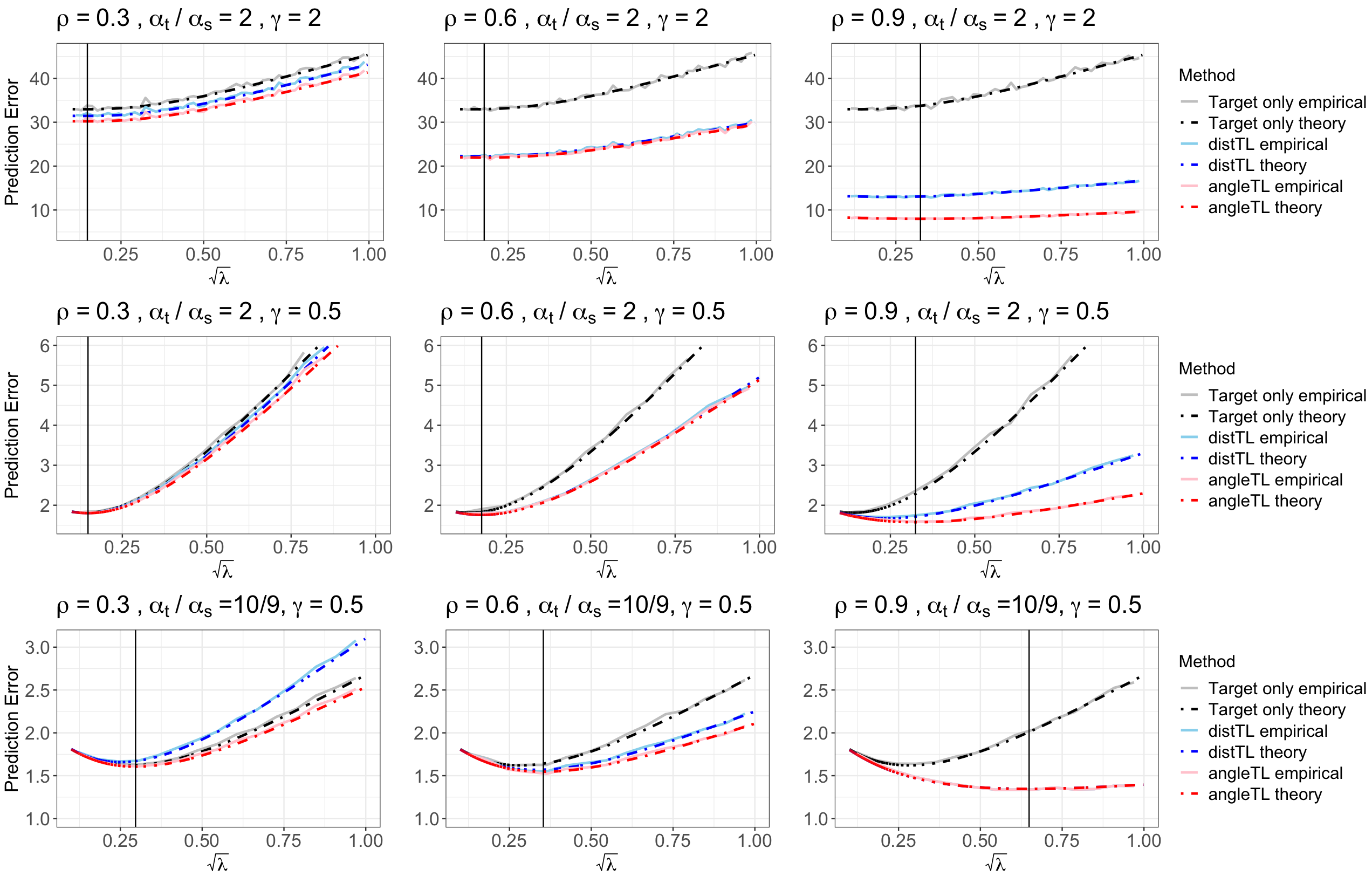} 
\caption{Empirical prediction error over 500 simulation replicates (solid curve) versus theoretical prediction error given in Equation (\ref{eq:noiseless}) (dashed curve). The vertical line shows the optimal $\sqrt{\lambda}$ in theory that leads to the minimum prediction error. We generate target and source estimates, $\bbeta$ and $\bw$, through multivariate Gaussian distribution, by varying their correlation $\rho \in \{0.3, 0.6, 0.9\}$ (in columns), signal strength ratio $\frac{\alpha_t}{\alpha_s} \in \{\frac{10}{9}, 2\}$ and dimension-to-sample ratio $\gamma=\frac{p}{n} \in \{2, \frac{1}{2}\}$ (in rows). We sequentially select 100 $\lambda$ values, and for each $\lambda$ we train on $50$ samples and test on 100 samples. We report the average test error over 500 simulations. Black: target-only model; blue: distTL; red: angleTL.} \label{plot_theorem1}
\end{figure}

Fig.\ref{plot_theorem1} compares the empirical risks and theoretical risks under the noiseless case. For all methods, the empirical risk aligns perfectly with the empirical risks, demonstrating the theoretical risk we obtained in Equation (\ref{eq:noiseless}) is precise in the noiseless case. As expected, angleTL is no worse than the target-only estimator and distTL across all settings. Moreover, the optimal $\lambda$ in theory (vertical line) precisely falls on the lowest point of the curve in all settings. Note that in the bottom right panel, we have $\rho \frac{\alpha_t}{\alpha_s}=1$ which represents the orthogonal case shown in panel C of Fig.\ref{similarity_measure}. We see that distTL reaches the same performance as angleTL. However, distTL can be worse than the target-only approach when $\rho\frac{\alpha_t}{\alpha_s}$ is far from $1$, e.g., the first column, indicating that it cannot prevent negative transfer. 

\begin{figure} 
\centering
\includegraphics[width=\textwidth]{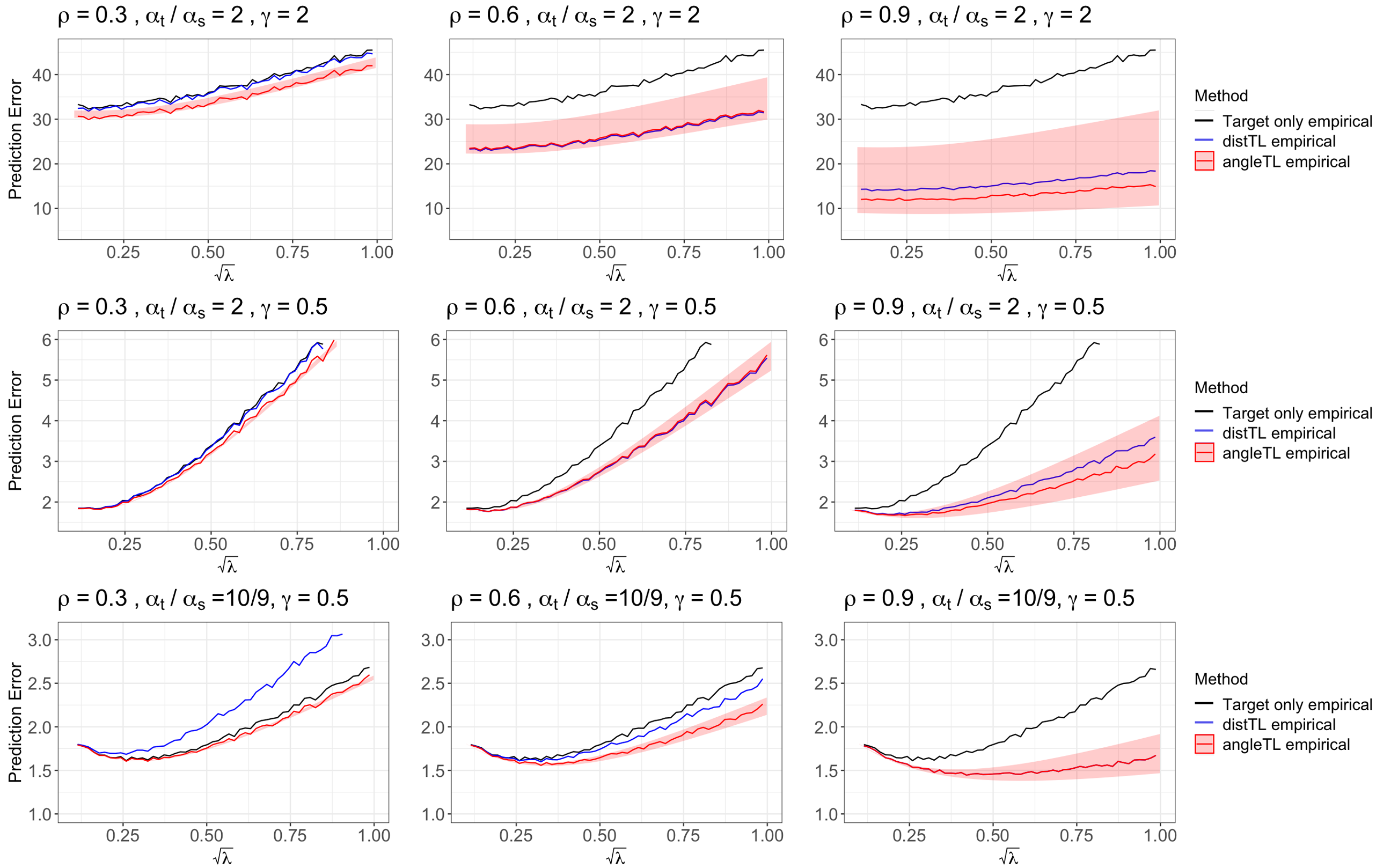} 
\caption{Empirical prediction error over 500 simulation replicates (solid curve) and theoretical prediction error bound of proposed method following Theorem \ref{withnoise} (filled area). We generate target and source estimates, $\bbeta$ and $\bw$, through multivariate normal distribution, by varying their correlation $\rho \in \{0.3, 0.6, 0.9\}$ (in columns), signal strength ratio $\frac{\alpha_t}{\alpha_s} \in \{\frac{10}{9}, 2\}$, and dimension-to-sample ratio $\gamma=\frac{p}{n} \in \{2, \frac{1}{2}\}$ (in rows). We add randomly generated and uniformly distributed noise $\bdelta$ to $\bw$. We sequentially take 100 $\lambda$ values, and for each $\lambda$ we train on $50$ samples and test on 100 samples. We report the average test error over 500 simulations. Black: target-only model; blue: distTL; red: angleTL.} \label{plot_theorem2}
\end{figure}

Similarly, we perform a simulation study to verify the predictive risks when there are estimation errors $\bdelta$. We generate $\bdelta \sim N(\textbf{0},\mathbf{\Sigma}_{\delta})$, where the covariance matrix $\mathbf{\Sigma}_{\delta}$ is generated by setting an exchangeable correlation of $0.1$ and the variances are uniformly generated between $0$ to $0.05$. We compute $C_L$ and $C_U$ using $\mathbf{\Sigma}_{\delta}$ and obtain the theoretical upper and lower bounds for the predictive risk of angleTL. Fig.\ref{plot_theorem2} shows the empirical predictive risk of the target-only estimator, distTL, and angleTL. The shaded area represents the theoretical lower and upper bounds of the predictive risk of angleTL obtained in Theorem \ref{withnoise}. We can see that the empirical risk of angleTL falls between the theoretical risk bounds. The width of the risk bounds increases with $\gamma$ and $\rho$. Across all scenarios, the upper bounds are lower than the empirical risk of the target-only approach, which shows that angleTL can protect against negative transfer.

\subsection{Evaluate the predictive performance across different settings}

\subsubsection{Single source study} \label{sim_setup_I}
We first evaluate the empirical predictive performance of angleTL in the case with one source study. To mimic the practical situation, we generate data for both the source and target populations based on the corresponding model parameters $\bw$ and $\bbeta$ where the corresponding entries are generated from Equation (\ref{simugen}). We set the sample sizes to be $N = 5,000$ for the source population, and $n = 50$ for the target population. The dimension of the model is set to $p \in \{25, 50, 100\}$. For the source population, we generate the predictors $\tilde \bX \sim N(\textbf{0},\mathbf{\tilde \Sigma})$, where the covariance matrix $\mathbf{\tilde \Sigma}$ is set to have variances of $1$ and exchangeable correlation of $0.2$, and we generate the outcome variable $\tilde \bY \sim N(\tilde \bX\bw, 0.5\mathbf{I}_N)$. For the target population, we generate $\bX \sim N(\textbf{0},\mathbf{\Sigma})$, where $\mathbf{\Sigma}$ contains variances of $1$ and exchangeable correlation of $0.1$, and we generate the outcome variable $\bY \sim N(\bX\bbeta, 0.5\mathbf{I}_n)$.

We apply an ordinary least square approach on the generated source data to obtain $\hat\bw$, which is then used together with the target data. We compare the predictive performance of angleTL to three state-of-art methods: (i) target-only: target-only estimator through Equation (\ref{ridge_tradition}); (ii) source-only: directly apply the source estimates on the testing data; (iii) distTL: distance-based transfer learning approach via Equation (\ref{distance-based_solution}). The predictive performance is evaluated by the root mean squared error (RMSE) calculated from an independent testing data set of size $200$, following the same data generating mechanism as the target data. The code for fitting target-only estimator,  distTL, and angleTL is the same with different choices of $\eta$ and $\lambda$ selected by a three-fold cross-validation, which ensures comparability. Specifically, when running target-only ridge regression, we set $\eta=0$ and only search $\lambda$ from a grid containing 100 points randomly selected between 0.0001 and 0.5. When running distTL, we set $\eta=\lambda$ and search $\lambda$ from a grid containing 100 points randomly selected between 0.0001 and 0.5. When running angleTL, $\eta$ and $\lambda$ are chosen from a two-dimensional grid, with $\lambda$ ranges between 0.0001 and 0.5 while $\eta$ ranges between 0.0001 and 0.2.

\begin{figure} 
\centering
\includegraphics[width=\textwidth]{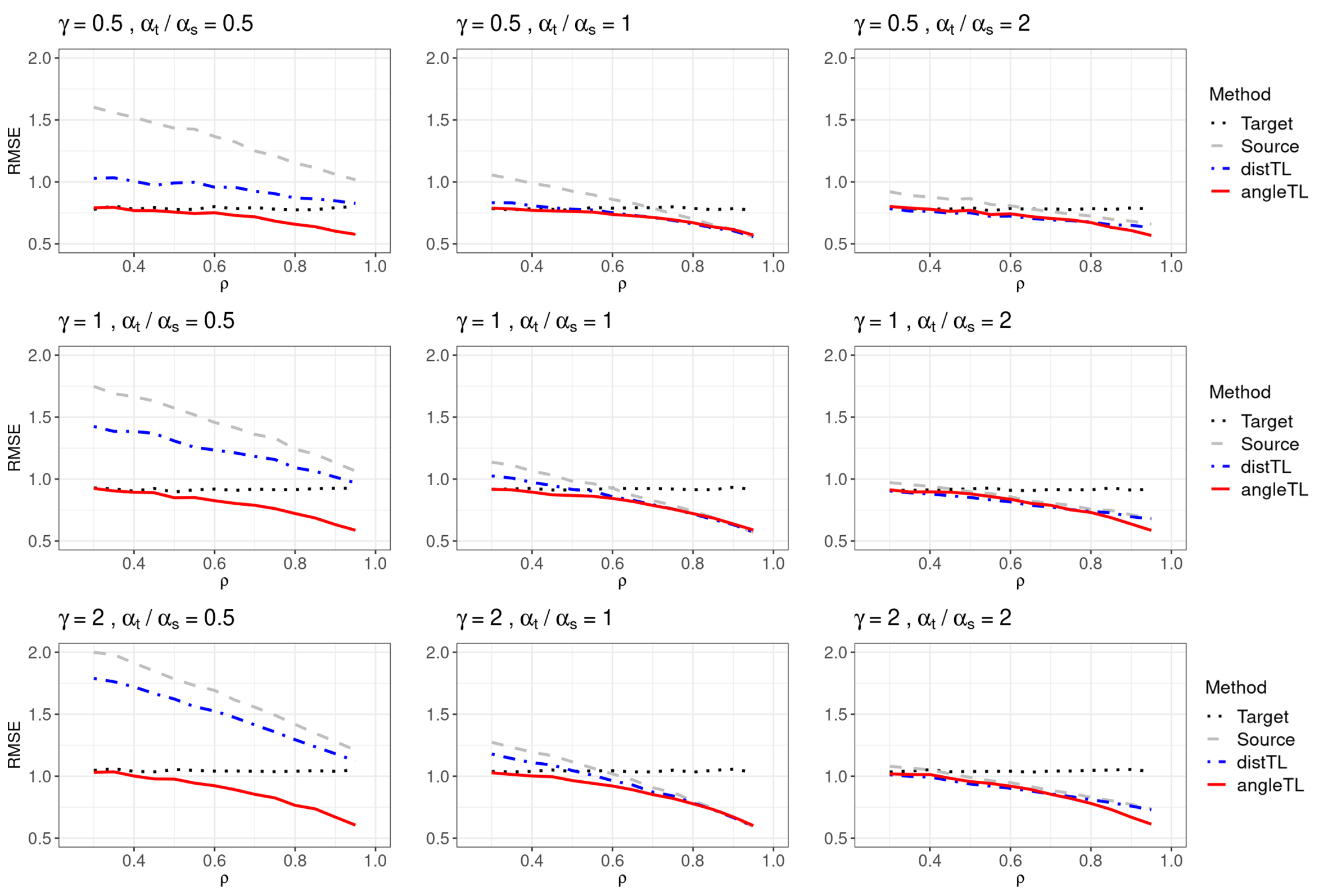} 
\caption{Root mean squared error (RMSE) of predicted outcome over 200 simulations. In each panel, we vary the correlation $\rho$ between the target and source estimates from 0.3 to 0.95. We vary $\gamma=\frac{p}{n} \in \{\frac{1}{2}, 1, 2\}$ (in rows) and the signal strength ratio $\frac{\alpha_t}{\alpha_s} \in \{\frac{1}{2}, 1, 2\}$ (in columns). Black: target-only model; grey: source-only model; blue: distTL; red: angleTL.} \label{plot_sim_1}
\end{figure}

In Fig.\ref{plot_sim_1}, each panel of RMSE is summarized from 200 independent simulations, where lower value represents better prediction accuracy. The pattern aligns with what we see in Fig. \ref{plot_theorem1}, where the performance of angleTL improves over increasing correlation between the target and the source estimates increases. When $\frac{\alpha_t}{\alpha_s}=2$, we see that distTL overlaps with angleTL around $\rho=\frac{1}{2}$ where $\rho \frac{\alpha_t}{\alpha_s}$ is close to $1$. When $\frac{\alpha_t}{\alpha_s}=\frac{1}{2}$, for all $\rho\in (0, 1)$, since $\rho \frac{\alpha_t}{\alpha_s}<\frac{1}{2}$, distTL shows consistently higher RMSE than angleTL; it also underperforms the target-only estimator, indicating that distTL fails to capture the similarity between the target and the source estimates and fails to prevent negative transfer. For a given $\gamma=\frac{p}{n}$, the performance of angleTL is stable across different values of $\frac{\alpha_t}{\alpha_s}$. For a given $\frac{\alpha_t}{\alpha_s}$, the source model helpful when the dimension of the model is higher.

\subsubsection{Multiple source studies}

We also consider the situation where we have multiple source studies as introduced in Section \ref{multi_source}. The target sample size is set to $n=100$ and each of the five source populations is of size $N=5,000$. We compare the results of the proposed multi-source estimators, angleTL-multi1 and angleTL-multi2, with the target-only estimator and the proposed single-source angleTL applying on the best single-source estimates. In angleTL1, we reserve 30\% of the target data (30 samples) for aggregation. 

\begin{figure} 
\centering
\includegraphics[width=\textwidth]{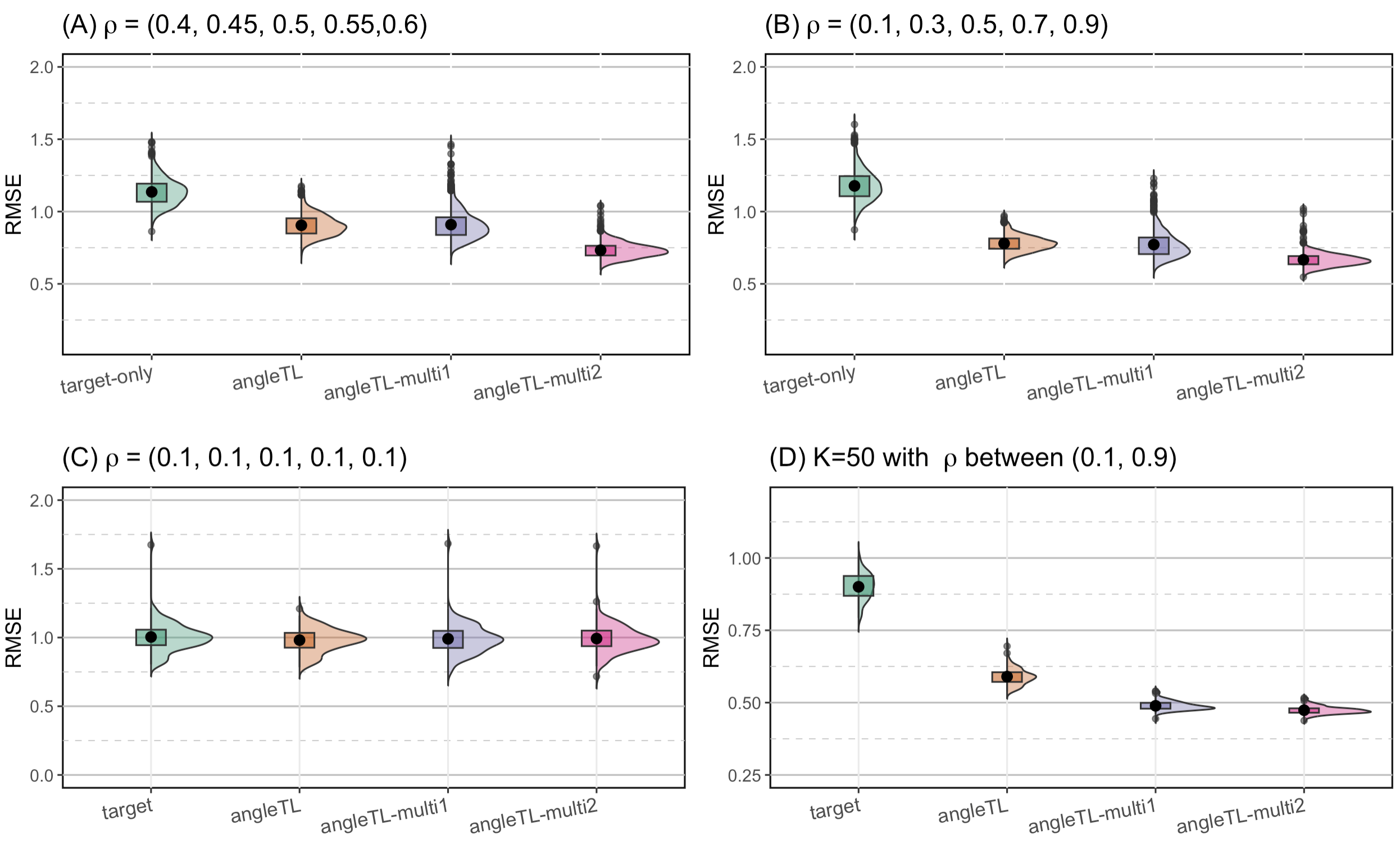} 
\caption{Root mean squared error (RMSE) of predicted outcome over 1,000 simulations comparing target only estimates (target-only), proposed angleTL when transferring from the best single model with the smallest $\rho$  (angleTL), and proposed angleTL combined with Algorithm 1 (angleTL-multi1) and and proposed angleTL combined with Algorithm 2 (angleTL-multi2). 
    Panels A, B, and C correspond to configurations with correlation $\rho=(\rho_1,\rho_2,\rho_3,\rho_4,\rho_5)$ being $(0.4,0.45,0.5,0.55,0.6)$  $(0.1,0.3,0.5,0.7,0.9)$,and $(0.1,0.1,0.1,0.1,0.1)$, respectively. Panel D corresponds to the setting where $K=50$ and correlations range from $0.1$ to $0.9$.}   \label{plot_sim_2}
\end{figure}

Fig.\ref{plot_sim_2} (A) shows the case where all source estimates have similar correlation as the target estimates between 0.4 to 0.6 and Fig.\ref{plot_sim_2} (B) shows the case where some of the sources are much more helpful than others, with correlations ranging from 0.1 to 0.9. In the former case, the proposed angleTL-multi2 outperforms angleTL which only transfers from the best single source model ($\hat\bw_5$ with $\rho_5=0.6$) and angleTL-multi1 with supervised aggregation. The later case shows similar pattern, where angleTL with the best single source ($\hat\bw_5$ with $\rho_5=0.9$) and angleTL-multi1 show improved performance than target-only and source-only models, slightly worse than angleTL-multi2. {In Fig.\ref{plot_sim_2} (C),  all five source vectors were very different from the target with $\rho = 0.1$. AngleTL shows similar performances as the target-only estimator,  without borrowing much information from source models with low similarity to the target. However, this demonstrates the robust performance of angleTL in terms of  preventing  negative transfer from irrelevant source models. } {When dealing with a large number of source models, we conduct a simulation study with $K = 50$, with $\rho$ ranging from 0.1 to 0.9 (Fig. \ref{plot_sim_2} (D)). Specifically, we randomly assign $\rho$ from Uniform$(0.1, 0.9)$ for 50 sources. We observe that angleTL-multi1 and angleTL-multi2 perform nearly the same and are better than target and single-source angleTL.}

\section{\Large Application to predict low-density lipoprotein} \label{realdata}

We apply angleTL, angleTL-multi1, and angleTL-multi2 to predict low-density lipoprotein (LDL) cholesterol, a blood biomarker that plays a crucial role in determining cardiovascular disease \citep{kwiterovich2000metabolic}, in self-reported White population using data from Mass General Brigham Biobank (MGBB) \citep{Partners_Biobank}. MGBB is a research database launched in 2021, containing around eighty thousands DNA samples before data quality control, yet still a relatively limited data size compared to other population-based biobanks such as the United Kingdom Biobank (UKB) \citep{ukbiobank2015cathie}. In this application, our goal is to build a prediction model for LDL using basic demographic and clinical risk factors, combined with top single nucleotide polymorphisms (SNPs), where our target population is the White population at MGBB. We consider a total of eight source models, including three models trained at UKB (LDL, Apolipoprotein B-100 [ApoB], and triglycerides [Tri]) and five models for LDL trained at the five participating sites at the electronic MEdical Records and GEnomics (eMERGE) Network \citep{gottesman2013electronic}.

In the MGBB database, we extracted data from 5,600 self-reported White participants. We extract up to the latest five LDL measures and use the average LDL as the outcome. The predictors include demographics such as age and gender, a binary indicator flagging the use of antihyperlipidemic medications, as well as related genetic variants selected from a clumping and thresholding procedure with linkage disequilibrium  and p-values of a genome-wide associate study (GWAS) using all the training samples from MGBB data \citep{choi2020tutorial}.  A total of 5,306 SNPs are selected with $R^2$ threshold set to 0.6 and p-value threshold set to 0.0005 using PLINK 2.0 \citep{chang2015second}.

We use UKB as one source dataset, from which we identify 409,031 samples with at least one measure of Tri, 408,602 samples with at least one measure of LDL, and 407,366 samples with at least one measure of Apo-B. To reduce the potentially inaccurate reporting ancestry, we selected those whose self-reported ancestry information agrees with the computed PC-based ancestry prediction \citep{zhangPC2020}.
    
Among the five source datasets in eMERGE, we identified a total of 16,723 samples with at least one LDL measure. We extract the same set of genetic, medication, and demographic variables from UKB and eMERGE, and fit eight source models using linear regression. A summary of basic information of the target and eight source datasets can be found in Table ~\ref{real_data_table}. Additional details regarding the data processing procedures can be found in Appendix 4.1 of the Supplementary Material.

\begin{table}
\caption{Basic information of the target and eight source datasets.} 
\resizebox{\textwidth}{!}{
\centering
\begin{tabular}{|l|ll|l|l|l|l|}
\hline
& \multicolumn{2}{l|}{\textbf{Biobank}} & \textbf{Outcome} & \textbf{Sample size} & \textbf{Mean age (SD)} & \textbf{Male (\%)}\\ \hline
Target  & \multicolumn{2}{l|}{MGBB} & LDL & 5,600 & 64.7 (15.3) & 2,959 (52.8)   \\ \hline
\multirow{8}{*}{Source} & \multicolumn{2}{l|}{\multirow{3}{*}{UKB}} & LDL & 408,602 & 69.8 (8.0) & 187,607 (45.9) \\ \cline{4-7} 
& \multicolumn{2}{l|}{}  & Tri  & 409,031  & 69.8 (8.0)& 187,821 (45.9) \\ \cline{4-7} 
& \multicolumn{2}{l|}{}  & Apo-B   & 407,366  & 69.8 (8.0)    & 186,763 (45.8) \\ \cline{2-7} 
& \multicolumn{1}{l|}{\multirow{5}{*}{eMERGE}} & Marshfield Clinic (Marsh) & LDL & 4,551  & 70.8 (9.7) & 1,496 (39.2)  \\ \cline{3-7} 
& \multicolumn{1}{l|}{} & University of Washington (UW)   & LDL & 618   & 80.9 (0.5) & 295 (54.0) \\ \cline{3-7} 
& \multicolumn{1}{l|}{} & Mayo Clinic (Mayo)  & LDL  & 3,185   & 71.3 (10.1) & 1,512 (56.7)\\ \cline{3-7} 
& \multicolumn{1}{l|}{}  & Northwestern University (NWU)  & LDL   & 3,037   & 53.4 (15.2)  & 287 (20.0) \\ \cline{3-7} 
& \multicolumn{1}{l|}{}  & Mount Sinai Hospital (MtSinai)  & LDL & 5,332  & 60.4 (13.1) & 363 (69.1) \\ \hline
\end{tabular}}
\label{real_data_table}
\end{table}

We compare the performance of the target-only estimator, the source-only estimator, distTL, and angleTL. We randomly split the target MGBB data into training and testing with a ratio of 10:1. For angleTL-multi1, which needs validation data for aggregation, we further split 10\% from the training data. All methods are trained on the training data, and the performance is measured by the $R^2$ (variance explained by the model) evaluated on the testing data. This process is repeated 100 times to account for the sampling variability. We employed ridge regression to calculate the target-only estimator and the majority of the source estimators, with the exception of those calculated from the UKB. Given that the total sample size in UKB exceeds 400,000, we opted for an ordinary least square linear regression due to computational efficiency.

\begin{figure}
\centering
\includegraphics[width=.8\textwidth]{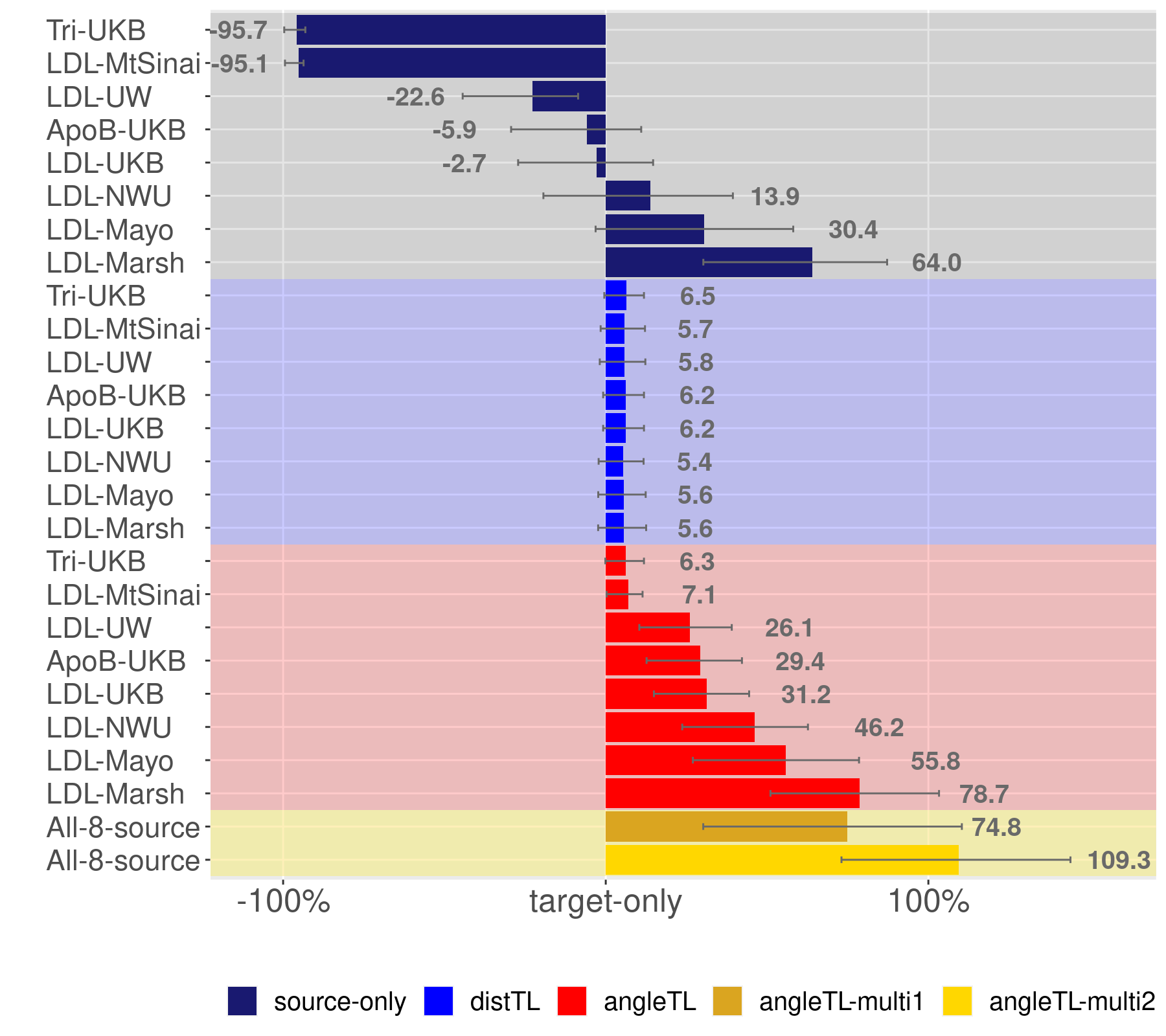} 
\caption{Mean ($\rm 1^{st}$-$\rm 3^{rd}$ quantile) percentage change of $R^2$ compared to the target-only model. The dark blue bars with grey background contains source-only estimates from UKB and eMERGE; the light blue bars with light blue background contains distTL; the red bars with red background includes the proposed single-source angleTL; and the last two yellow rows show angleTL-multi1 and angleTL-multi2, respectively. Tri-UKB: triglycerides from UKB; ApoB: Apolipoprotein B-100; LDL: low-density lipoprotein; MtSinai=Mount Sinai Hospital, UW=University of Washington, NWU=Northwestern University, Mayo=Mayo Clinic, and Marsh=Marshfield Clinic.} \label{lipids}
\end{figure}

In Fig.\ref{lipids}, we present the relative performance of each method compared to the benchmark target-only model. With 100 replications, we report the mean, the first and the third quantiles of the percentage improvement of $R^2$. Among all the source-only models (dark blue bars with grey background), the model trained at UKB using triglycerides as the outcome has the worse performance, and the model trained at Marshfield Clinic has the highest performance, where the average improvement is around 64\%. When transferring each source model to train the target model using distTL (light blue bars with light blue background), all the single-source distTL have similar performance, around 6\% improvement. When applying single-source angleTL (red bars with red background), we observe that the performance is uniformly better than the corresponding source-only models, and angleTL also outperforms distTL. In the last two rows, we see that angleTL-multi1 has comparable performance as the best single source, LDL-Marsh, while angleTL-multi2 has the largest average improvement of 109.3\%. {To assess the generalizability of the conclusion when the target population changes, we treat each of the sites in eMERGE as the target population, and obtain similar results where angleTL achieves the best performance in nearly all cases. More details can be found in Appendix 4.2 of the Supplementary Material.}

{In this real application involving real data, the source models are derived using information from diverse study cohorts with varied outcomes. This results in heterogeneity on two levels: the cohort level and the outcome level, when comparing each source model to the target model. For instance, UKB is a population-based biobank, representing a comparatively healthier population than those from hospital-based cohorts collected at the eMERGE sites. Furthermore, the outcomes under consideration extend beyond LDL to include correlated traits such as triglycerides and Apolipoprotein B-100. These traits might possess overlapping genetic architectures with LDL. However, the influence of each genetic variant on the traits may differ. Such heterogeneity requires more adaptive adjustment methods than the $L_2$ constraint offered by distTL. This might explain why angleTL outperforms distTL in this particular application.}

\section{Discussion} \label{discussion}
We propose angleTL, a flexible transfer learning framework that leverages the concordance of model parameters across populations. Compared to several benchmark methods, including the target-only model, the source-only models, and the distance-based transfer learning method, angleTL is shown to have improved performance both empirically and numerically. Under the setting with a single source model, we identify factors that influence the predictive performance of angleTL, including (1) the similarity between model parameters measured by the $\sin\Theta$ distance, (2) the signal strengths of the source and the target model, (3) the ratio between model dimensionality and the target sample size, (4) the estimation error of the source estimator, and (5) the level of the residual noise, which provide useful practical guidance for understanding when a source population can be helpful. With multiple source models, we propose to aggregate the source models first before applying angleTL, which we also provide theoretical justifications of the aggregation algorithms. Like ridge regression, angleTL is easy-to-implement as it enjoys the advantage of having a closed-form solution. It does not require individual-level data from the source, and can prevent negative transfer. Our simulation and real data application demonstrate the validity and feasibility of angleTL across a wide range of settings.

One limitation of angleTL is that it requires the source and the target models to include the same covariates, whereas, in practice, source models might include covariates that are not measured at the target data or vice versa. There are existing methods focusing on incorporating source models that use a subset of covariates in the target model. For example, \cite{Chatterjee2016CML} proposed a constrained maximized likelihood approach where the external data follows the same distribution as the internal data. Accounting for heterogeneous data, methods such as \cite{estes2018empirical,kundu2019generalized,zhang2020generalized,gu2020meta,taylor2022data} are proposed, which all considered low-dimensional models and their ability to handle high-dimensional data requires further investigation. A more detailed discussion on related work, including different empirical approaches, can be found in \citet{han2022discussion}.

It is an interesting future direction to extend angleTL to the setting where the source study uses a subset of the covariates in the target study. Methods proposed in related work such as \cite{taylor2022data} can be considered to address this, where we can first map the unique variables to a space orthogonal to the shared variables, and then include them in the model while using the angle-based regularization to borrow information.  

While we employ a linear model to illustrate the concept of angleTL, it is important to note that the penalty term $\lambda\|\bbeta\|_2^2-2\eta\hat\bw^\top\bbeta$ can be seamlessly integrated with the objective functions of various other problems. For instance, in addition to the regression problem discussed in our paper, we can apply this approach to a classification problem. In this scenario, our objective is to classify an outcome variable $y$ based on a covariate vector $\bx$, using a classification rule denoted as $f(\bx^\top\bbeta)$. When we aim to learn this classification rule for the target population, we have the flexibility to select an appropriate objective function. Furthermore, we can incorporate source estimators $\hat{\bw}_k$ derived from different populations using various models, such as logistic regression, support vector machines \citep{cortes1995support}, and linear or quadratic discrimination analysis \citep{hastie2009elements}, into the target model through the angleTL penalty (an additional simulation study to demonstrate the performance of angleTL in a classification problem can be found in Appendix 3 of  Supplementary Material).

While the proposed penalization is motivated by an  angle-based similarity measure  between the source and target model parameters, in a classification problem, it might be useful to consider an alternative similarity defined by the fitted outcomes. More specifically, we can consider the similarity measure defined as $\sin\Theta(X\bbeta,X\bw)$ or a rank-based measure
$
\sum_{i\neq j} \mathbb{I}(X_i^\top \bbeta > X_j^\top \bw, X_i^\top \bbeta > X_j^\top \bw)$, which is closely related to the maximum rank correlation problem discussed in \cite{han1987non, wang2007note, stephanou2021sequential, shin2021exact}.

\section*{Acknowledgement}
This work is supported by the National Institute of General Medical Sciences (NIGM) R01GM148494.

\bibliographystyle{rss}
\bibliography{refs.bib}

\end{document}